\documentclass[12pt,reqno]{amsart}
\usepackage{natbib}
\usepackage{booktabs}       % professional-quality tables
\usepackage{amsfonts}       % blackboard math symbols
\usepackage{nicefrac}       % compact symbols for 1/2, etc.
\usepackage{graphicx}
\usepackage{subcaption}
\usepackage{longtable}
\usepackage{enumitem,kantlipsum}
\usepackage{rotating}
\usepackage{array}
\usepackage{mathtools}

\usepackage{wrapfig}
\usepackage{url,lineno,microtype,subcaption}
\usepackage[onehalfspacing]{setspace}
\usepackage{blindtext}
\usepackage[ruled,vlined]{algorithm2e}
\usepackage{cite}
\usepackage{bm}

\usepackage{amsmath,amsthm}
\theoremstyle{definition}

\usepackage[dvipsnames]{xcolor}
\makeatletter
 \def\@textbottom{\vskip \z@ \@plus 1pt}
 \let\@texttop\relax
\makeatother
\usepackage[margin=3cm]{geometry}
\usepackage[final]{hyperref} % adds hyper links inside the generated pdf file
\hypersetup{
	colorlinks=true,       % false: boxed links; true: colored links
	linkcolor=blue,        % color of internal links
	citecolor=blue,        % color of links to bibliography
	filecolor=magenta,     % color of file links
	urlcolor=blue         
}
\usepackage[foot]{amsaddr}
\title[Causal Functional Connectivity in Alzheimer's Disease]{Application of Time-Aware PC Algorithm to compute Causal Functional Connectivity in Alzheimer's Disease from fMRI data}
\author[Biswas]{Rahul Biswas$^\ast$}
\address{$^\ast$Department of Electrical and Computer Engineering, University of Washington, Seattle, WA, 98195, USA. \textit{E-mail for correspondence:} \textnormal{\texttt{rbiswas1@uw.edu}.}}

\author[Sripada]{SuryaNarayana Sripada$^{\ast\ast}$}
\address{$^{\ast\ast}$Center for Research on Science and Consciousness, Redmond, WA, 98052, USA.}

\begin{document}
\maketitle

\begin{abstract}
\emph{Functional Connectivity} between brain regions is known to be altered in Alzheimer's disease, and promises to be a biomarker for early diagnosis of the disease. While several approaches for functional connectivity obtain an un-directed network representing stochastic associations (correlations) between brain regions, association does not necessarily imply causation. In contrast, \emph{Causal Functional Connectivity} is more informative, providing a directed network representing causal relationships between brain regions. In this paper, we obtained the causal functional connectome for the whole brain from recordings of resting-state functional magnetic resonance imaging (rs-fMRI) for subjects from three clinical groups: cognitively normal, mild cognitive impairment, and Alzheimer's disease. We applied the recently developed \emph{Time-aware PC} (TPC) algorithm to infer the causal functional connectome for the whole brain. TPC supports model-free estimation of whole brain causal functional connectivity based on directed graphical modeling in a time series setting. We then perform an exploratory analysis to identify the causal brain connections between brain regions which have altered strengths between pairs of subject groups, and over the three subject groups, based on edge-wise p-values from statistical tests. We used the altered causal brain connections thus obtained to compile a comprehensive list of brain regions impacted by Alzheimer's disease according to the current data set. The brain regions thus identified are found to be in agreement with literature on brain regions impacted by Alzheimer's disease, published by researchers from clinical/medical institutions. %The obtained brain regions are in agreement with existing literature published by researchers from clinical/medical institutions thus validating the approach. %We then discuss the soundness and completeness of the results and the potential for using  causal functional connectivity obtained using this methodology as a basis for the prognosis and diagnosis of Alzheimer's disease.

\end{abstract}

\section{Introduction}
Alzheimer's disease (AD) is the most common age-related progressive neurodegenerative disorder. It typically begins with a preclinical phase and advances through mild cognitive impairment (MCI) to clinically significant AD, which is a form of dementia \citep{querfurth2010alzheimer}. Despite significant efforts to identify biomarkers for AD, it still relies on clinical diagnosis, and early and accurate prediction of the disease remains limited \citep{laske2015innovative, li2019deep}. Abnormal resting-state functional connectivity (FC) between brain regions has been observed as early as two decades before brain atrophy and the emergence of AD symptoms \citep{ashraf2015cortical,nakamura2017early}. Therefore, resting-state FC can potentially determine the relative risk of developing AD \citep{brier2014functional,sheline2013resting}.

Resting-state functional magnetic resonance imaging (rs-fMRI) records the blood-oxygen-level-dependent (BOLD) signal from different brain regions while individuals are awake and not engaged in any specific task. The BOLD signal is popularly used to infer functional connectivity between brain regions partly due to the advantage that BOLD signal provides high spatial resolution \citep{yamasaki2012understanding,sporns2013human,liu2015multimodal,xue2019distinct}.

Functional connectivity refers to the stochastic relationship between brain regions with respect to their activity over time. Popularly, functional connectivity involves measuring statistical association between signals from different brain regions. The statistical association measures are either pairwise associations between pairs of brain regions such as Pearson's correlation, or multivariate i.e. incorporating multi-regional interactions such as undirected graphical models \citep{biswas2021statistical}. Detailed technical explanations of functional connectivity in fMRI can be found in \citet{chen2017methods,keilholz2017time,scarapicchia2018resting}. The findings from studies using FC \citep{wang2007altered,kim2016distinctive}, and meta-analyses \citep{jacobs2013meta,li2015toward,badhwar2017resting} indicate a decrease in connectivity in several brain regions in relation to Alzheimer's disease (AD), such as the hippocampus and posterior cingulate cortex. These regions play a role in memory and attentional processing. On the other hand, some studies have found an increase in connectivity within brain regions in early stages of AD and MCI \citep{gour2014functional,bozzali2015impact, hillary2017injured}. This is a well known phenomenon, where increase in FC between certain brain regions occurs when the communication between other brain regions is impaired. Such hyperconnectivity has been interpreted as a compensatory mechanism where alternative paths within the brain's network are recruited \citep{hillary2017injured,marek2022frontoparietal,oldham2019development}. 

In contrast to associative FC, causal FC represents functional connectivity between brain regions more informatively by a directed graph, with nodes as the brain regions, directed edges between nodes indicating causal relationships between the brain regions, and weights of the directed edges quantifying the strength of the corresponding causal relationship \citep{spirtes2000causation}. However, functional connectomics studies in general, and in relation to fMRI from Alzheimer's disease in particular, have predominantly used associative measures of FC \citep{reid2019advancing}. There are a few studies focusing on alterations in CFC in relation to Alzheimer's disease \citep{rytsar2011inhibition, khatri2021diagnosis}, however this area is largely unexplored. This is partly due to the lack of methods that can infer the CFC in a desirable manner as explained next.

Several properties are desirable in the context of causal modeling of functional connectivity \citep{biswas2021statistical, smith2011network}. Specifically, the CFC should represent causality while free of limiting assumptions such as linearity of interactions. In addition, since the activity of brain regions are related over time, such temporal relationships should be incorporated in defining causal relationships in neural activity. The estimation of CFC should be computationally feasible for the whole brain functional connectivity, instead of limiting to a smaller brain network. It is also desirable to capture beyond-pairwise multi-regional cause and effect interactions between brain regions. Furthermore, since the BOLD signal occurs and is sampled at a temporal resolution that is far slower than the neuronal activity, thereby causal effects often appear as contemporaneous \citep{granger1969investigating, smith2011network}. Therefore, the causal model in fMRI data should support contemporaneous interactions between brain regions.

Among the methods for finding CFC, \emph{Dynamic Causal Model} (DCM) requires a mechanistic biological model and compares different model hypotheses based on evidence from data, and is unsuitable for estimating the CFC of the whole brain \citep{friston2003dynamic, smith2011network}. On the other hand, Granger Causality typically assumes a vector auto-regressive linear model for activity of brain regions over time, and it tells whether a regions's past is predictive of another's future \citep{granger2001essays}. Furthermore, GC does not include contemporaneous interactions. This is a drawback since fMRI data often consists of contemporaneous interactions \citep{smith2011network}. In contrast, \emph{Directed Graphical Modeling} (DGM) has the advantage that it does not require the specification of a parametric equation of the neural activity over time, it is predictive of the consequence of interventions, and supports estimation of whole brain CFC. Furthermore, the approach inherently goes beyond pairwise interactions to include multiregional interactions between brain regions, along with estimation of the cause and effect of such interactions. The \emph{Time-aware PC} (TPC) algorithm is a recent method for computing the CFC based on DGM in a time series setting \citep{biswas2022statistical2}. In addition, TPC also incorporates contemporaneous interactions among brain regions. A detailed comparative analysis of approaches to find causal functional connectivity is provided in \citet{biswas2021statistical, biswas2022statistical2}. With the development of methodologies such as Time-aware PC, it would be possible to infer the whole brain CFC with the aforementioned desirable properties.

In this paper, we apply the TPC algorithm to infer the causal functional connectivity between brain regions from resting-state fMRI data. By applying the algorithm to the fMRI of subjects, we estimate the subject-specific CFC for all subjects in the dataset. It is noteworthy that different subjects are in different clinical categories: Cognitively Normal (CN), Mild Cognitive Impairment (MCI), and Alzheimer's Disease (AD). To identify which causal connections differ between brain regions across pairs of clinical categories, we utilize Welch's t-test, comparing the weights of causal functional connections of subjects for a pair of clinical categories. This analysis reveals the p-values of causal links between brain regions to exhibit differences between subjects in distinct clinical categories, such as with cognitively normal vs. Alzheimer's disease. Additionally, we employ a Kruskal-Wallis H-test, which is a non-parametric version of ANOVA test, to find p-values of causal links to exhibit differences across subjects of the three clinical categories. These links provide insights into the causal connectivity connections that are relevant in exhibiting differences among the three clinical categories. We then compile a comprehensive list of brain regions impacted by Alzheimer's disease based on the altered causal links obtained from the current dataset. Notably, the obtained brain regions are consistent with existing literature, with each such publication being a report from a team involving a clinical setting and at least one medical expert, thereby validating the approach. %We then discuss the soundness and completeness of the results and the potential for using  causal functional connectivity obtained using this methodology as a basis for the prognosis and diagnosis of Alzheimer's disease.

\section{Materials and Methods}
\subsection{Participants}
The resting fMRI and demographic data were downloaded from the Harvard Dataverse (\href{https://doi.org/10.7910/DVN/29352}{https://doi.org/10.7910/DVN/29352}) \citep{harvarddata}. A total of 30 subjects were included in the study: 10 subjects who are cognitively normal (CN), 10 subjects with mild cognitive impairment (MCI), and 10 subjects with Alzheimer's disease (AD).

In the experiments, general cognitive evaluation of subjects was obtained using the Mini-Mental State Examination (MMSE) \citep{harvarddata}. The subjects were age-matched (ANOVA test: $F = 1.5, p > 0.2$) and gender-matched (chi-square test: $\chi^2 = 1.9, p > 0.3$), although subjects with MCI or AD were less educated than subjects in CN group (t-tests: AD vs CN, $t = -4.0, p < 0.001$; MCI vs CN, $t = -2.3, p < 0.05$). As expected, MMSE scores had a significant difference between all pairs of groups (t-tests: AD vs CN, $t = -6.5, p < 0.001$; MCI vs CN, $t = -4.6, p < 0.001$; MCI vs AD: $t = 3.1, p < 0.05$).

Table \ref{tab:data} includes a summary of the participants' demographic and medical information.

\begin{table}[t]
    \centering
\caption{Subject demographic information summary}\label{tab:data}
% \bgroup
% \def\arraystretch{1.6}
    \begin{tabular}{cccc}
    \hline
         & CN & MCI & AD\\
         \hline
        $n$ & $10$ & $10$ & $10$ \\
        Sex (M/F) & $7/3$ & $6/4$ & $4/6$ \\
        Age (years) & $66.0\pm 9.6$ & $70.7 \pm 7.1$ & $72.3\pm 8.3$ \\
        Education (years) & $14.5 \pm 3.0$ & $11.1 \pm 3.5$ & $8.6 \pm 3.6$ \\
        MMSE & $29.30\pm 0.67$  & $25.8\pm 2.3$ & $21.5\pm 3.7$\\
        %CDR & $1.24\pm 2.88$  & $4.18\pm 1.07$ \\
        \hline
    \end{tabular}
    \vspace{2mm}
    
Entries represent the mean $\pm$ S.D.
\end{table}

\subsection{Image Acquisition}
The acquisition of fMRI images was performed using a Siemens Magnetom Allegra scanner. The fMRI images were obtained using an echo planar imaging (EPI) sequence at a field strength of $3.0$ Tesla, with a repetition time (TR) of $2.08$ seconds, an echo time (TE) of $30$ milliseconds, and a flip angle of $70$ degrees. The matrix size was $64 \times 64$ pixels, there were 32 axial slices parallel to AC-PC plane, in plane resolution was $3\times 3$ mm$^2$, slice thickness was 2.5 mm. Resting scans lasted for 7 mins and 20 secs for a total of $220$ volumes during which subjects were instructed to keep their eyes closed, to not think of anything in particular and to refrain from falling asleep.

\subsection{fMRI Preprocessing}
The fMRI pre-processing steps were carried out using the CONN toolbox version 21a, which utilizes the Statistical Parametric Mapping (SPM12), both of which are MATLAB-based cross-platform software \citep{nieto2021conn,friston1994statistical}. We used the default preprocessing pipeline in CONN, consisting of the following steps in order: functional realignment and unwarp (subject motion estimation and correction), functional centring to (0,0,0) coordinates (translation), slice-time correction with interleaved slice order, outlier identification using Artifact Detection and Removal Tool (ART), segmentation into gray matter, white matter and cerebrospinal fluid tissue, and direct normalization into standard Montreal Neurological Institute (MNI) brain space, and lastly, smoothing using spatial convolution with a Gaussian kernel of 8mm full width half maximum. This pipeline was followed by detrending, and bandpass filtering (0.001-0.1 Hz) to remove low-frequency scanner drift and physiological noise of the fMRI images. The first four time points have been filtered out to remove any artifact.

For the extraction of Regions-Of-Interest (ROIs), the automated anatomical labeling (AAL) atlas was utilized on the preprocessed rs-fMRI dataset \citep{tzourio2002automated}. The list of all regions in AAL atlas is provided in Appendix \ref{appen:aal} along with their abbreviated, short, and full region names. This specific parcellation method has been demonstrated to be optimal for studying the functional connectivity between brain regions \citep{arslan2018human}. The voxels within each ROI were averaged, resulting in a time series for each ROI. %The resulting time series consisted of $216$ timepoints, with a repetition time (TR) of $2.08$ seconds, totaling $449.28$ seconds.

\subsection{Inference of causal functional connectivity: Time-aware PC algorithm}
\begin{figure}[t!]
    \centering
    \href{https://drive.google.com/file/d/1XiJY9c74TtfbDYphwwKVtMjPEo02_hhE/view?usp=sharing}{\includegraphics[width = 0.9\textwidth]{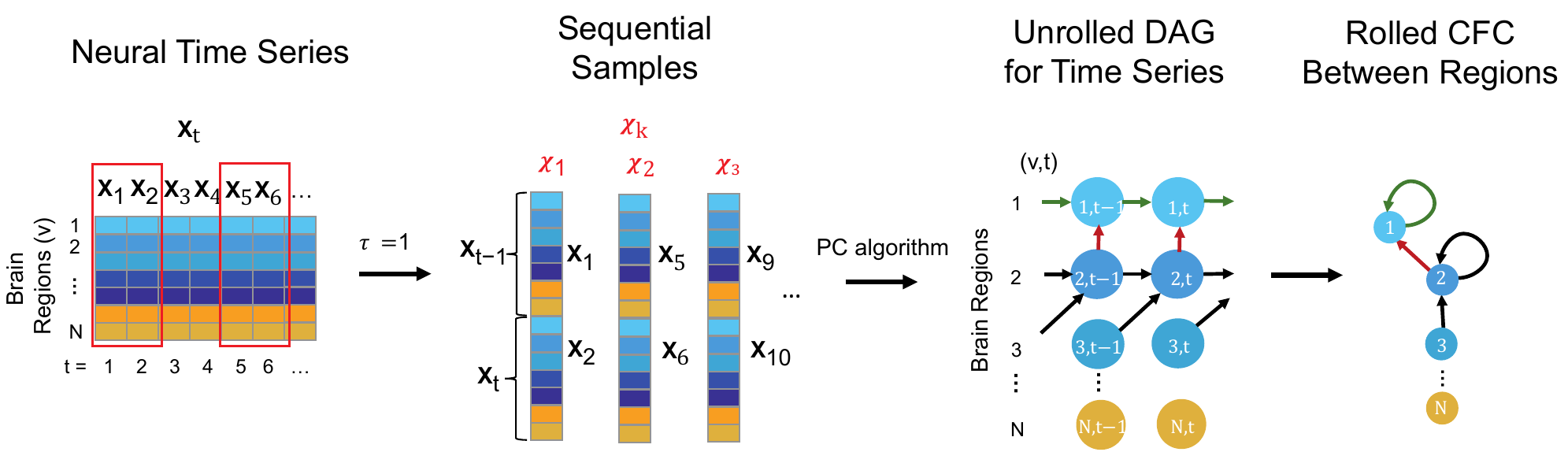}}
    \caption{Steps conveying the concept of the TPC algorithm to infer CFC from observed neural time series data: First the neural time series is transformed to form sequential samples with a maximum time delay of interaction, $\tau$ (here $\tau$ = 1). Then, Peter-Clark (PC) algorithm is applied on the sequential samples to obtain the unrolled DAG satisfying the Directed Markov Property. Finally the unrolled DAG is transformed to obtain the Rolled CFC between regions.}
    \label{fig:tpc}
\end{figure}

The Time-aware PC (TPC) Algorithm finds causal functional connectivity between brain regions from time series based on Directed Graphical Models (DGM) \citep{spirtes2000causation, pearl2009causality, biswas2021statistical,biswas2022statistical2, biswas2022consistent}. While traditional DGM is applicable to static data, TPC extends the applicability of DGM to CFC inference in time series by firstly implementing the Directed Markov Property (DMP) to model causal spatial and temporal interactions in the time series by an unrolled Directed Acyclic Graph (DAG) of the time series. The unrolled DAG consists of nodes $(v,t)$, for region of interest $v$ and time $t$, and edge $(v_1,t_1)\rightarrow (v_2,t_2)$ reflecting causal interaction from the BOLD signal in region $v_1$ at time $t_1$ to the BOLD signal in region $v_2$ at time $t_2$. The estimation of the unrolled DAG is carried out by first transforming the time series into sequential variables with a maximum time delay of interaction $\tau$, and then applying the Peter-Clark (PC) algorithm to infer the unrolled DAG based on the sequential variables \citep{kalisch2007estimating}. TPC then rolls the DAG back to obtain the CFC graph between the regions of interest (see Figure \ref{fig:tpc}) \citep{biswas2022statistical2}. We consider $\tau = 1$ for our analyses, which would include interactions of the BOLD signal between regions of interest with a maximum time delay of 2.08 s, the TR of the fMRI acquisition.

The CFC outcome of this methodology is interpretable in the following manner: An edge from region $i\rightarrow j$ in the CFC estimate represents significant causal interaction from brain region $i$ at preceding times to region $j$ at following times. The model and the approach are non-parametric, meaning that it does not require the specification of a parametric dynamical equation for neural activity. The method captures beyond-pairwise multivariate interactions between brain regions. It also supports estimation of the CFC for the whole brain in a computationally feasible manner. It also allows for time delays in interactions between the brain units as well as the presence of feedback-loops. Furthermore, it has been shown that if the neural activity obeys an arbitrary dynamical process, the model outcome of TPC is consistent with respect to the causal relationships implied by the dynamical process and is predictive of counterfactual queries such as ablation or modulation \citep{biswas2022statistical2}.

It is noteworthy that implementing the DMP on the unrolled DAG to model causal relationships over time enables contemporaneous interactions e.g. from region $u$ to region $v$ at time $t$ \citep{biswas2022statistical2}. Such contemporaneous interactions are represented by the edge $(u,t)\rightarrow (v,t)$ in the unrolled DAG, and presence of such an edge in the unrolled DAG would be reflected as an edge $u\rightarrow v$ in the Rolled CFC outcome. Such contemporaneous interactions are especially relevant in fMRI due to the relatively slow temporal resolution of the BOLD signal compared to the underlying neural activity \citep{smith2011network}.

\subsection{Group-wise Comparisons of Estimated CFC}
Using the subject-specific CFC estimated by TPC algorithm, we perform further statistical tests. We use Welch's t-test to obtain p-values of connections to exhibit greater or less weight in one disease stage compared to another disease stage \citep{yuen1974two}. We use the Kruskal-Wallis H-test, which is a non-parametric version of the ANOVA test, to obtain p-values of connections to exhibit unequal weight in either of the four disease stages \citep{kruskal1952use}.

\section{Results}
\subsection{Subject-specific Causal Functional Connectivity}

\begin{figure}[t!]
    \centering
    \href{https://drive.google.com/file/d/1PESbwDAfwjLpT2RTwHBYCfkOrOKNN4ic/view?usp=sharing}{\includegraphics[width = 0.95\textwidth]{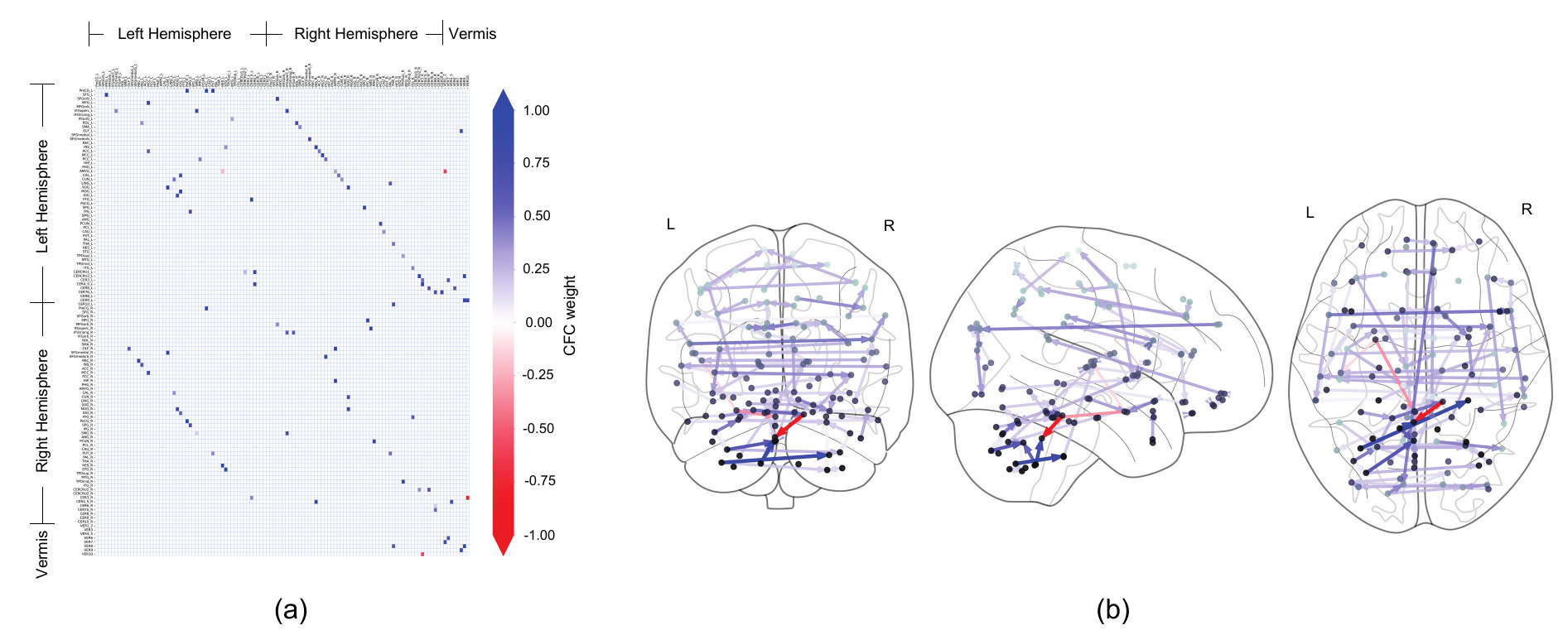}}
    \caption{CFC outcome of an example subject who is cognitively normal. The CFC is obtained by TPC algorithm. (a) The CFC is represented by a matrix, whose entry $(i,j)$ represents the connection of region $i\rightarrow j$. (b) The CFC is visualized with graph edges on the Frontal, Axial and Lateral brain maps (left to right). The nodes correspond to brain region centers, ranging from superficial (light gray) to deeper (darker gray) regions, in the AAL brain atlas.}
    \label{fig:cfcexample}
\end{figure}

\begin{figure}[h!]
    \centering
    %\begin{subfigure}{\textwidth}
    \href{https://drive.google.com/file/d/1T9HxdIIblqNOTtf5CDgi2LFbSLtEf_-U/view?usp=sharing}{\includegraphics[width = 0.95\textwidth]{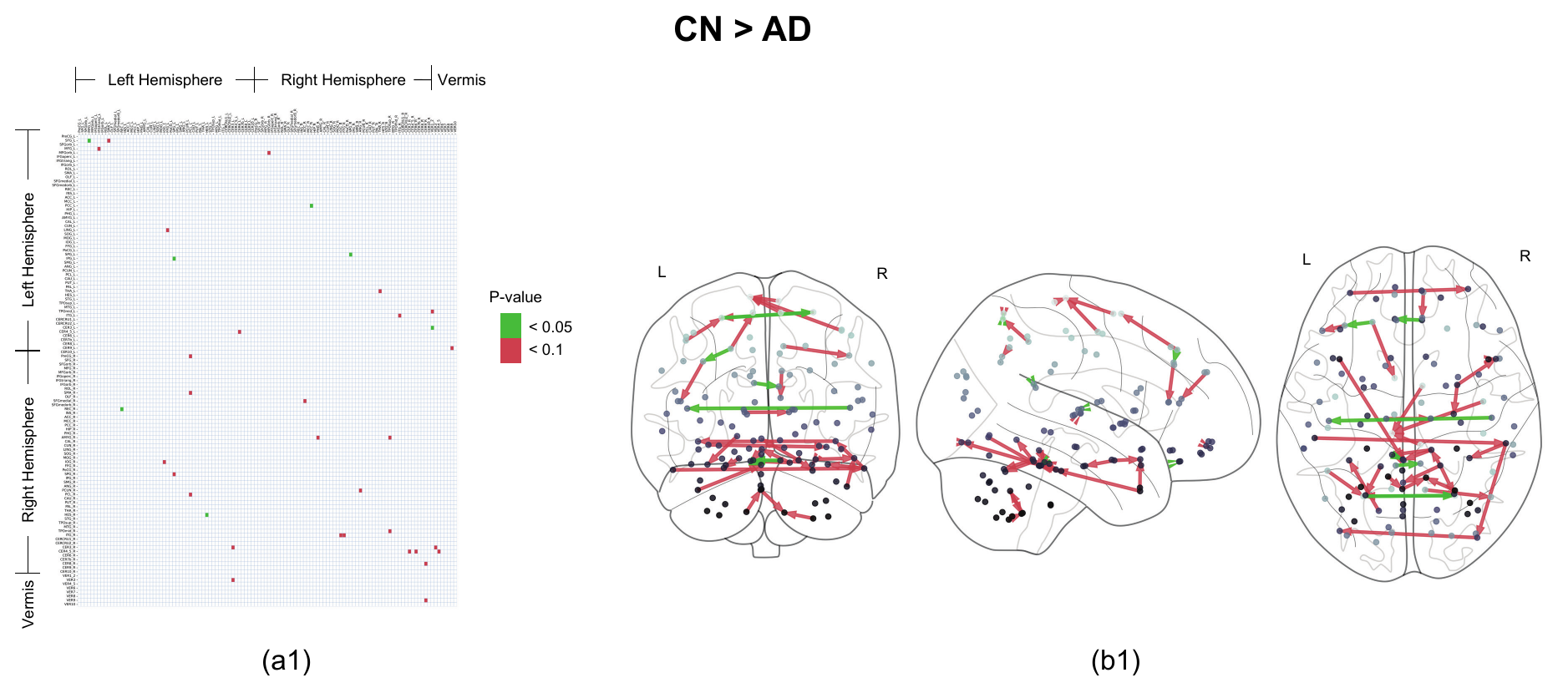}}\\
    \href{https://drive.google.com/file/d/1-smROT0-TXLNvm5-YmElIklf2DOEqG3b/view?usp=sharing}{\includegraphics[width = 0.95\textwidth]{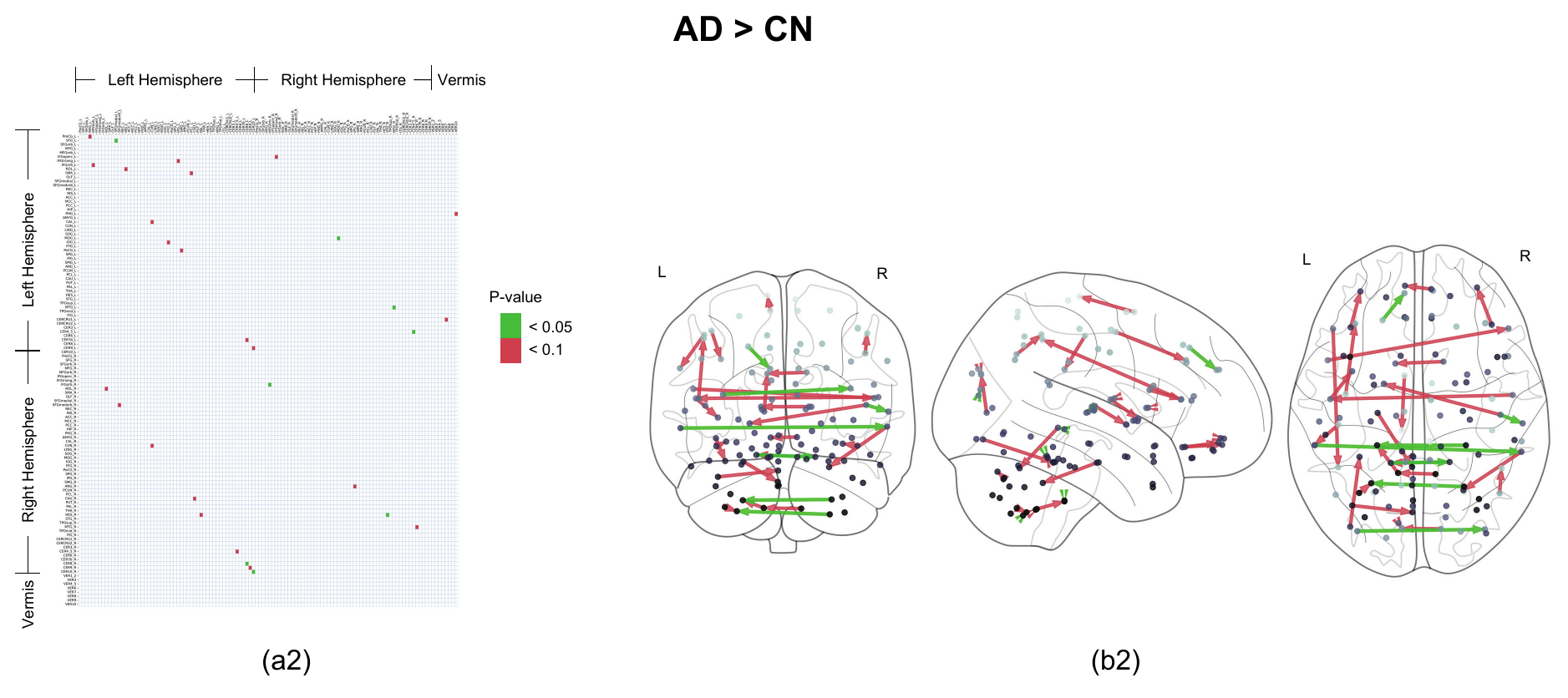}}\\
    \href{https://drive.google.com/file/d/1-vMBZ1dpo90ZlQJUgGVLjtzqOp5-wQAL/view?usp=sharing}{\includegraphics[width = 0.95\textwidth]{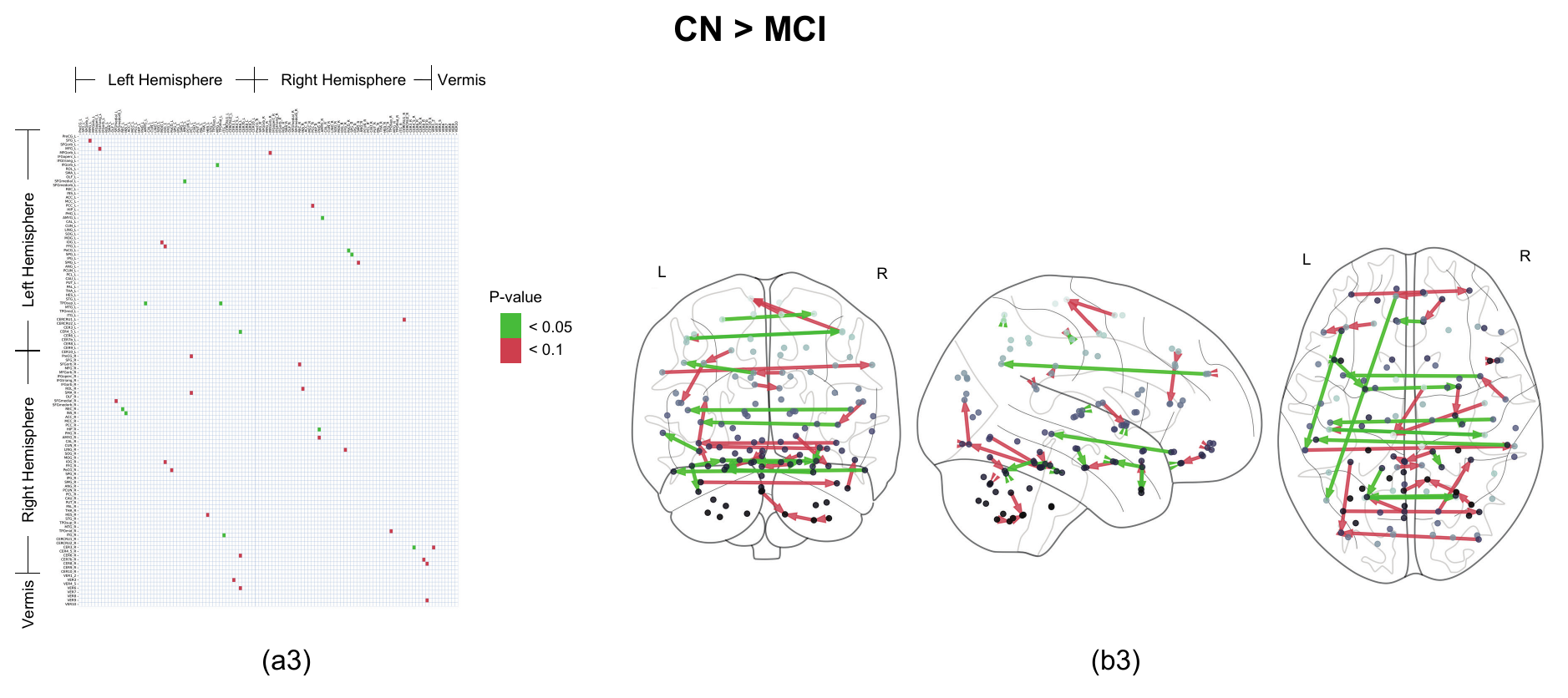}}
    \caption{Causal functional connections with p-values for altered weights in pariwise comparisons between groups less than 0.05 (green) and less than 0.1 (red), as obtained by Welch's t-test. (a1)-(a3). The connections are represented in matrix format with a non-zero entry in $(i,j)$ corresponding to the edge $i\rightarrow j$. (b1)-(b3). The connections are represented by graph edges on the brain schematics.}% (top) and level $0.025$ (bottom).}
    \label{fig:pwcomp0.05}
\end{figure}
\begin{figure}[h!]
    \ContinuedFloat
    \centering
    \href{https://drive.google.com/file/d/1mzrBhWfa-l4cs40XAdBX2SxuqdUAUICh/view?usp=sharing}{\includegraphics[width = 0.95\textwidth]{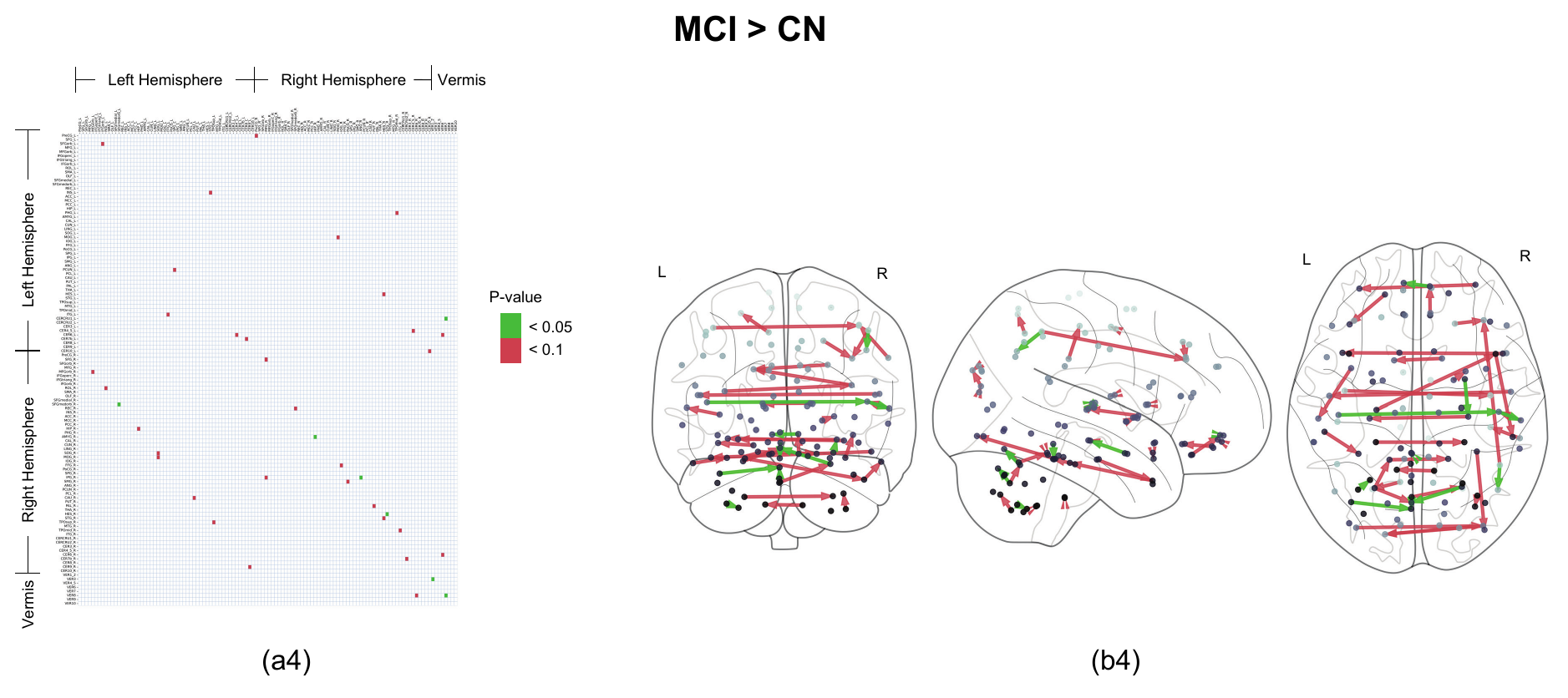}}\\
    \href{https://drive.google.com/file/d/1jnpMCf0q8rJgC4tQsjkoU7twVQv6QLL1/view?usp=sharing}{\includegraphics[width = 0.95\textwidth]{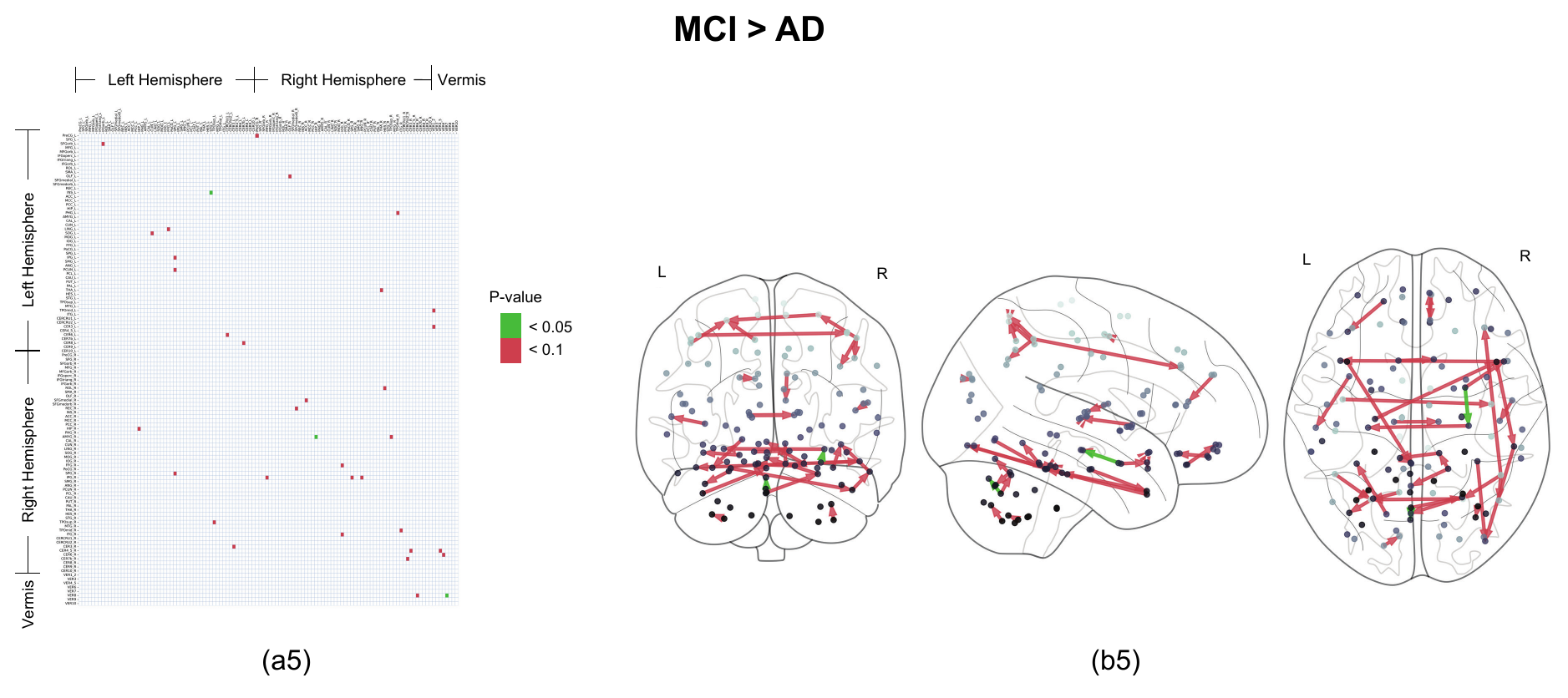}}\\
    \href{https://drive.google.com/file/d/1_2hyuz3kL-rIVTwJP7I5uyFWjRacz5kZ/view?usp=sharing}{\includegraphics[width = 0.95\textwidth]{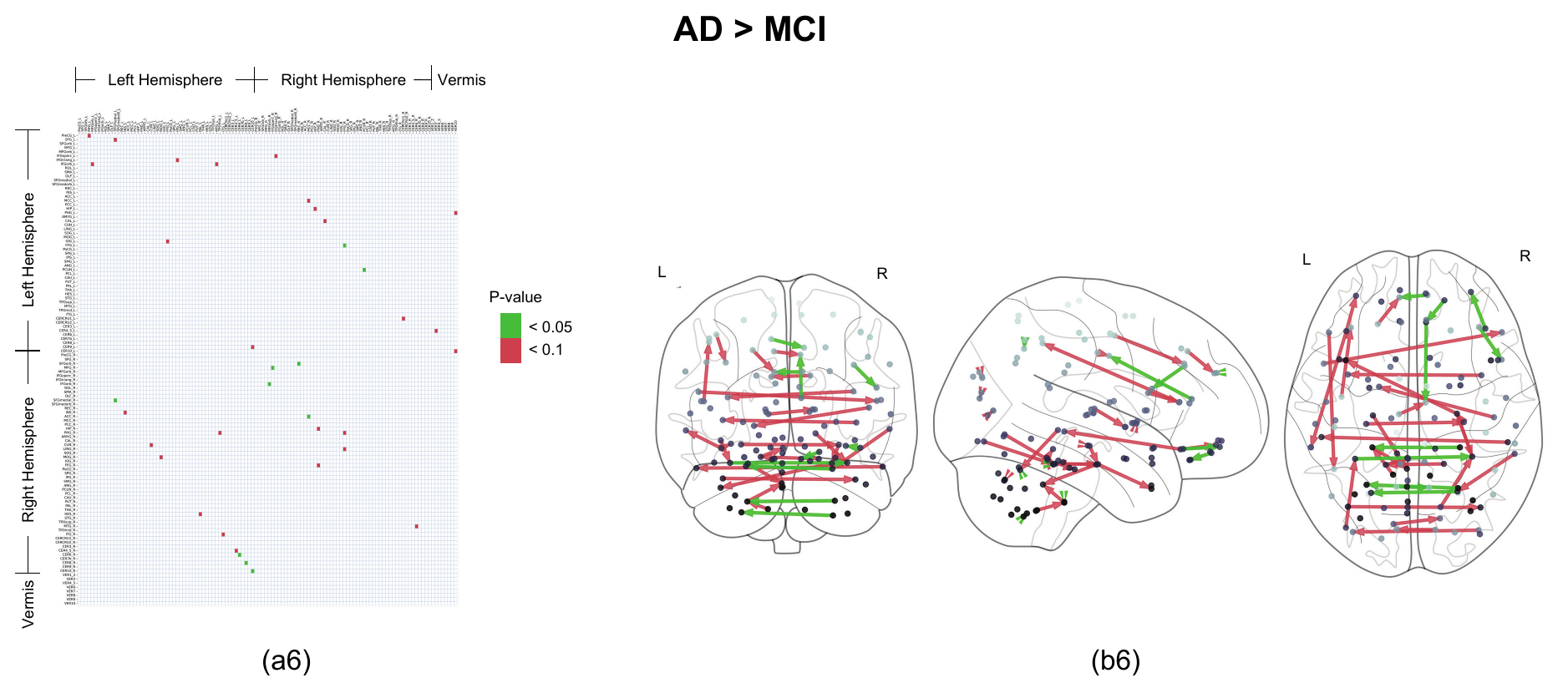}}
    \caption{(Cont.) Causal functional connections with p-values for altered weights in pariwise comparisons between groups less than 0.05 (green) and less than 0.1 (red), as obtained by Welch's t-test. (a1)-(a3). The connections are represented in matrix format with a non-zero entry in $(i,j)$ corresponding to the edge $i\rightarrow j$. (b1)-(b3). The connections are represented by graph edges on the brain schematics.}
    \label{fig:pwcomp0.05}
\end{figure}

Figure \ref{fig:cfcexample} shows the causal functional connectivity (CFC) estimated using TPC algorithm for an example subject in the cognitively normal group. In Figure \ref{fig:cfcexample}-a, the CFC is represented in the form of a matrix, whose entry $(i,j)$ indicates the presence of connectivity from region index $i\rightarrow j$, and the value at entry $(i,j)$ represents the weight of that causal connection. A positive value (blue) of the weight is indicative of excitatory influence whereas a negative value (red) is indicative of inhibitory influence. The diagonal of the matrix representing self-connections for regions has been filtered out. In Figure \ref{fig:cfcexample}-b, the CFC is represented by a directed graph overlayed on schematics of the brain. The schematics of the brain comprise 2-dimensional brain projections in the Frontal, Axial and Lateral planes. The nodes of the CFC graph correspond to centers of brain regions in the AAL atlas. The nodes are colored light to dark gray according to their depth in the brain, with light gray representing superficial and dark gray representing deeper brain regions. The causal functional connectivity graph provides a highly informative map of causal interactions between brain regions.

\subsection{Comparisons of causal functional connectivity over pairwise clinical categories}\label{sec:cfccomp}

Figure \ref{fig:pwcomp0.05} shows the CFC edges obtained by the TPC algorithm which have p-value less than 0.05 and less than 0.1 for altered (greater/lower) weight between subjects from pairs of disease stages, where the p-values are found by Welch's t-test. This provides insights into which brain connections are impacted by different stages of the disease: healthy, early and late mild cognitive impairment, and Alzheimer's disease. 

In Figure \ref{fig:pwcomp0.05}, the connections with lowest 5 p-values from AD$>$CN and AD$<$CN comparisons are: Lobule IV, V of cerebellar hemisphere Left $\rightarrow$ Lobule IV, V of cerebellar hemisphere Right; Superior frontal gyrus, dorsolateral Left $\rightarrow$ Superior frontal gyrus, medial Left; Middle occipital gyrus Left $\rightarrow$ Middle occipital gyrus Right; Gyrus rectus Right $\rightarrow$ Gyrus rectus Left, Heschl's gyrus Right $\rightarrow$ Heschl's gyrus Left.

\begin{figure}[t!]
\vspace{0.6in}
    \centering
    \href{https://drive.google.com/file/d/1eu-joYjSBWX3eMB2FG0cODw1qsmma-6d/view?usp=sharing}{\includegraphics[width = 0.95\textwidth]{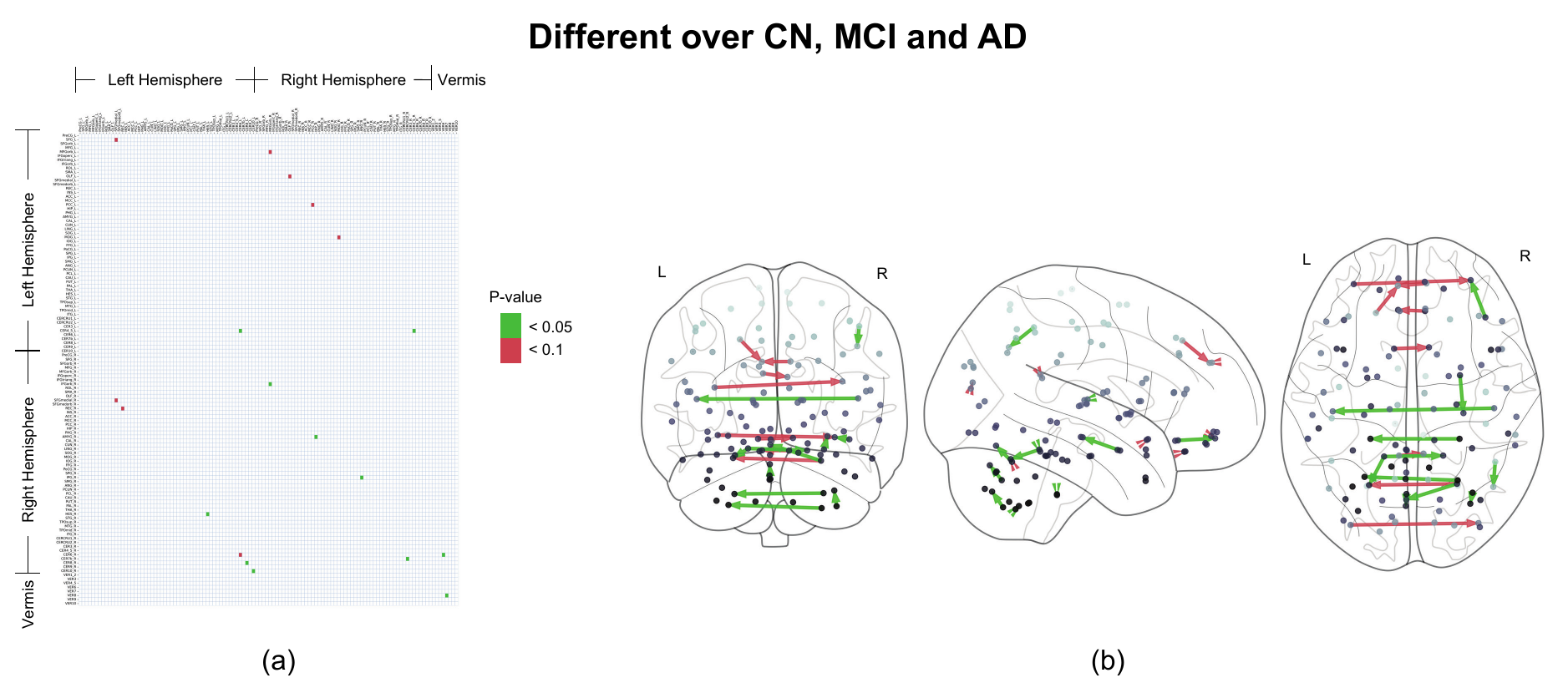}}
    \caption{Causal functional connections with p-values for altered weights between the three groups of CN, MCI and AD subjects less than 0.05 (green) and additional connections with p-values less than 0.1 (red), where the p-values are obtained by Kruskal-Wallis H-test. (a) The connections are represented in matrix format with a non-zero entry in $(i,j)$ corresponding to the edge $i\rightarrow j$. (b) The connections are represented by graph edges on the schematic of the brain.}
    \label{fig:anovacomp}
\end{figure}

\subsection{Multi-group comparison of causal functional connectivity}
Figure \ref{fig:anovacomp} shows the causal functional connections obtained by TPC, which have p-value less than 0.05 and less than 0.1 for difference across either of the four disease stages. Here the p-values are obtained by Kruskal-Wallis H-test. This sheds light on those connections which are impacted during the overall progression of the disease. Here, the connections with lowest 5 p-values are: Lobule VIIB of cerebellar hemisphere Right $\rightarrow$ Crus II of cerebellar hemisphere Right; Lobule IV, V of cerebellar hemisphere Left $\rightarrow$ Lobule IV, V of cerebellar hemisphere Right; Inferior frontal gyrus, pars orbitalis, Right $\rightarrow$ Middle frontal gyrus, pars orbitalis Right; Amygdala Right $\rightarrow$ Hippocampus Right; Lobule VI of cerebellar hemisphere Right $\rightarrow$ Lobule VI of vermis.
\subsection{Brain Regions with altered connections}
In Table \ref{tab:sigregions}, we list 9 brain regions (and 5 additional regions) that correspond to altered causal functional connections to or from them between subjects of CN, MCI and AD groups with edge-wise p-value less than 0.05 (less than 0.1). The subject-specific causal functional connectomes have been estimated using TPC algorithm (see Section \ref{sec:cfccomp}). The brain regions found are in agreement with existing publications cited in Table \ref{tab:sigregions}-right column.

%\restoregeometry

\newgeometry{top=2cm,left=2cm,right=2cm,bottom=2cm}
\begin{table}[h!]
\resizebox{\textwidth}{!}{
    \centering
\begin{tabular}{p{2.4in}p{2in}c}
\toprule
\textbf{Abbreviated Name} & \textbf{Region Name} & \textbf{Reported by} \\
\midrule\midrule
\textcolor{ForestGreen}{AMYG\_R} & Amygdala & \citet{vogt1990pathological,poulin2011amygdala}\\
\midrule
\textcolor{ForestGreen}{ANG\_L}, \textcolor{ForestGreen}{ANG\_R} & Angular gyrus &  \citet{benson1982angular,jagust2006brain} \\
\midrule
\textcolor{ForestGreen}{CAU\_R} & Caudate nucleus & \citet{barber2002volumetric,madsen20103d}\\
\midrule
\textcolor{ForestGreen}{CER4\_5\_L}, \textcolor{ForestGreen}{CER4\_5\_R}, \textcolor{ForestGreen}{CER6\_L}, \textcolor{ForestGreen}{CER6\_R}, \textcolor{ForestGreen}{CER7b\_R}, \textcolor{ForestGreen}{CER8\_L}, \textcolor{ForestGreen}{CER8\_R}, \textcolor{ForestGreen}{CER10\_L},
\textcolor{ForestGreen}{CER10\_R},
\textcolor{ForestGreen}{CERCRU2\_R},
\textcolor{Maroon}{CER6\_L},
\textcolor{Maroon}{CER6\_R} & Cerebellum & \citet{joachim1989diffuse,jacobs2018cerebellum} \\
\midrule
\textcolor{ForestGreen}{HES\_L}, \textcolor{ForestGreen}{HES\_R} & Heschl’s gyrus & \citet{hanggi2011volumes,dhanjal2013auditory} \\
\midrule
\textcolor{ForestGreen}{HIP\_R} & Hippocampus & \citet{ball1985new,boutet2014detection,rao2022hippocampus}\\
\midrule
\textcolor{ForestGreen}{IFGorb\_R} & Inferior frontal gyrus & \citet{eliasova2014non,cajanus2019association} \\
\midrule
\textcolor{ForestGreen}{IPG\_R} & Inferior parietal gyrus & \citet{de2015topography,van20157t}\\
\midrule
\textcolor{ForestGreen}{MFGorb\_R}, \textcolor{Maroon}{MFGorb\_L} & Middle frontal gyrus & \citet{neufang2011disconnection,zhou2013impaired}\\
\midrule
\textcolor{ForestGreen}{MTG\_R} & Middle temporal gyrus & \citet{busatto2003voxel, piras2019association} \\
\midrule

\textcolor{ForestGreen}{PCL\_L} & Paracentral lobule & \citet{garcia2013macular,yang2019study} \\
\midrule

\textcolor{ForestGreen}{VER6}, \textcolor{ForestGreen}{VER7}, \textcolor{ForestGreen}{VER8},  \textcolor{Maroon}{VER1\_2} & Vermis & \citet{sjobeck2001alzheimer,a2013dendritic}\\
\midrule

\textcolor{Maroon}{PCC\_L}, \textcolor{Maroon}{PCC\_R} & Cingulate gyrus & \citet{villain2008relationships,caminiti2020imaging, mascali2015}\\
\midrule
\textcolor{Maroon}{REC\_L}, \textcolor{Maroon}{REC\_R} & Gyrus rectus & \citet{molsa1987alzheimer,nochlin1993comparison, sheline2010amyloid}\\
\midrule
\textcolor{Maroon}{MOG\_L}, \textcolor{Maroon}{MOG\_R} & Middle occipital gyrus & \citet{golby2005memory,frings2015asymmetries}\\
\midrule
\textcolor{Maroon}{OLF\_L}, \textcolor{Maroon}{OLF\_R} & Olfactory cortex & \citet{reyes1993olfactory, wang2010olfactory}\\
\midrule
\textcolor{Maroon}{SFG\_L}, \textcolor{Maroon}{SFGmedial\_L},\newline \textcolor{Maroon}{SFGmedial\_R} & 
Superior frontal gyrus & \citet{brachova1993association,lue1996inflammation} \\
\bottomrule\bottomrule
\end{tabular}
}
\caption{Brain regions corresponding to altered causal functional connections, whose strength is altered between CN, MCI and AD subject groups with p-values less than 0.05 (green) and additionally with p-values less than 0.1 (red) based on Kruskal-Wallis H-test. The causal functional connections are obtained by TPC algorithm.}
\label{tab:sigregions}

\end{table}
\restoregeometry

\section{Discussion}
In this study, we have obtained the causal functional connectivity of the whole brain from resting state fMRI time series. We used the recently developed Time-aware PC (TPC) algorithm based on directed graphical modeling in time series, to compute the causal functional connectivity. In the dataset, the subjects belonged to three clinical categories: cognitively normal (CN), mild cognitive impairment (MCI) and Alzheimer's disease (AD). We performed group-wise comparisons of the subject-specific causal functional connectivity to identify which causal functional connections are altered between pairs of subject groups. The altered causal functional connections between CN and AD were used to obtain brain regions involved with such altered connections in AD. This resulted in the identification of 12 brain regions where causal functional connections to or from those regions are altered in Alzheimer's disease with p-value less than 0.05, and 5 additional regions with p-value less than 0.1.

It is noteworthy that while several studies have concluded decreased connectivity in MCI and AD compared to CN \citep{jacobs2013meta,li2015toward,badhwar2017resting}, prominent researchers have highlighted that MCI and early stages of AD can involve an increase in functional connectivity between brain regions. This increase occurs when the communication between specific brain regions is impaired. It has been interpreted as a compensatory mechanism where alternative paths within the brain's network are recruited \citep{hillary2017injured, marek2022frontoparietal,oldham2019development}. This explains the presence of causal functional connections estimated by TPC algorithm, whose weight in AD is greater compared to CN in addition to connections with weight in AD less than that in CN (see Figure \ref{fig:pwcomp0.05}). The following are causal functional connections found by TPC, that have edge-wise p-value less than 0.05 for strength in AD greater than that in CN, and are lowest 5 in p-value: Lobule IV, V of cerebellar hemisphere Left $\rightarrow$ Lobule IV, V of cerebellar hemisphere Right; Superior frontal gyrus, dorsolateral Left $\rightarrow$ Superior frontal gyrus, medial Left; Middle occipital gyrus Left $\rightarrow$ Middle occipital gyrus Right; Middle temporal gyrus Left $\rightarrow$ Middle temporal gyrus Right; Heschl's gyrus Right $\rightarrow$ Superior temporal gyrus Right.

In the short term, the augmentation of functional connectivity along alternative pathways exhibits efficiency and adaptability of the brain. However, it is imperative to acknowledge the susceptibility of these densely interconnected hubs to beta-amyloid deposition, which can elicit secondary damage through metabolic stress, ultimately culminating in system breakdown \citep{hillary2017injured}. Consequently, the initial state of hyperconnectivity observed in neurodegenerative disorders may gradually transition into hypoconnectivity among the engaged pathways, thereby contributing to cognitive decline as the disease advances \citep{marek2022frontoparietal}.

Based on the causal functional connectome outcome alone, this study has been able to identify many brain regions related to Alzheimer's disease, which have been reported across more than 30 different studies, using different feature extraction methods and advanced imaging technologies. Therefore, this study demonstrates the promise of a causal functional connectivity approach based on directed graphical models in time series and estimated by TPC algorithm. 

Although most of the regions linked to AD have been identified, certain regions such as the Thalamus \citep{vstepan2014cortical}, that have been also linked with AD has not been identified by the methodology. This can be due to lack of significance given the low sample size or choice of significance level. However, the study demonstrates the potential in using the methodology to a larger dataset. Using a larger dataset, the causal functional connectomics methodology can be used for 1) computation of a subject's causal functional connectome from their fMRI data, 2) identification of specific connections as biomarkers for Alzheimer's disease, and 3) using biomarker connections for the early prognosis and diagnosis of Alzheimer's disease.

In this paper, we have demonstrated the following: (a) Application of the TPC algorithm to compute whole-brain CFC for each subject, (b) Interpretation of CFC in the context of AD using domain (neuropathological) knowledge, and (c) Exploratory analysis for edge-wise differences and corresponding brain regions with altered connectivity in subjects from pairs of clinical categories (CN, MCI and AD), and among the three clinical categories. The findings are consistent with published medical literature. In summary, our results show the promise of computing the whole-brain CFC from fMRI data using the TPC algorithm to gain prognostic and diagnostic insights.

\bibliographystyle{apalike}
\bibliography{main}

\appendix
\section{Automated Anatomical Labeling (AAL) Atlas}\label{appen:aal}

The regions in the AAL atlas along with their abbreviated, short and full names are listed in Table \ref{tab:aal}. 

\begin{longtable}{rlll}
\caption{Names of regions in the AAL Atlas}\label{tab:aal}\\
\toprule
 No &    Abbr. Name &            Short Name &                                           Full Region Name \\
\midrule
  1 &     PreCG\_L &         Precentral\_L &                              Precentral gyrus Left \\
  2 &     PreCG\_R &         Precentral\_R &                             Precentral gyrus Right \\
  3 &       SFG\_L &        Frontal\_Sup\_L &          Superior frontal gyrus, dorsolateral Left \\
  4 &       SFG\_R &        Frontal\_Sup\_R &         Superior frontal gyrus, dorsolateral Right \\
  5 &    SFGorb\_L &    Frontal\_Sup\_Orb\_L &        Superior frontal gyrus, pars orbitalis Left \\
  6 &    SFGorb\_R &    Frontal\_Sup\_Orb\_R &       Superior frontal gyrus, pars orbitalis Right \\
  7 &       MFG\_L &        Frontal\_Mid\_L &                          Middle frontal gyrus Left \\
  8 &       MFG\_R &        Frontal\_Mid\_R &                         Middle frontal gyrus Right \\
  9 &    MFGorb\_L &    Frontal\_Mid\_Orb\_L &          Middle frontal gyrus, pars orbitalis Left \\
 10 &    MFGorb\_R &    Frontal\_Mid\_Orb\_R &         Middle frontal gyrus, pars orbitalis Right \\
 11 &  IFGoperc\_L &   Frontal\_Inf\_Oper\_L &        Inferior frontal gyrus, opercular part Left \\
 12 &  IFGoperc\_R &   Frontal\_Inf\_Oper\_R &       Inferior frontal gyrus, opercular part Right \\
 13 & IFGtriang\_L &    Frontal\_Inf\_Tri\_L &       Inferior frontal gyrus, triangular part Left \\
 14 & IFGtriang\_R &    Frontal\_Inf\_Tri\_R &      Inferior frontal gyrus, triangular part Right \\
 15 &    IFGorb\_L &    Frontal\_Inf\_Orb\_L &       Inferior frontal gyrus, pars orbitalis, Left \\
 16 &    IFGorb\_R &    Frontal\_Inf\_Orb\_R &      Inferior frontal gyrus, pars orbitalis, Right \\
 17 &       ROL\_L &      Rolandic\_Oper\_L &                            Rolandic operculum Left \\
 18 &       ROL\_R &      Rolandic\_Oper\_R &                           Rolandic operculum Right \\
 19 &       SMA\_L &    Supp\_Motor\_Area\_L &                      Supplementary motor area Left \\
 20 &       SMA\_R &    Supp\_Motor\_Area\_R &                     Supplementary motor area Right \\
 21 &       OLF\_L &          Olfactory\_L &                              Olfactory cortex Left \\
 22 &       OLF\_R &          Olfactory\_R &                             Olfactory cortex Right \\
 23 & SFGmedial\_L & Frontal\_Sup\_Medial\_L &                Superior frontal gyrus, medial Left \\
 24 & SFGmedial\_R & Frontal\_Sup\_Medial\_R &               Superior frontal gyrus, medial Right \\
 25 & SFGmedorb\_L &    Frontal\_Med\_Orb\_L &        Superior frontal gyrus, medial orbital Left \\
 26 & SFGmedorb\_R &    Frontal\_Med\_Orb\_R &       Superior frontal gyrus, medial orbital Right \\
 27 &       REC\_L &             Rectus\_L &                                  Gyrus rectus Left \\
 28 &       REC\_R &             Rectus\_R &                                 Gyrus rectus Right \\
 29 &       INS\_L &             Insula\_L &                                        Insula Left \\
 30 &       INS\_R &             Insula\_R &                                       Insula Right \\
 31 &       ACC\_L &       Cingulum\_Ant\_L &       Anterior cingulate \& paracingulate gyri Left \\
 32 &       ACC\_R &       Cingulum\_Ant\_R &      Anterior cingulate \& paracingulate gyri Right \\
 33 &       MCC\_L &       Cingulum\_Mid\_L &         Middle cingulate \& paracingulate gyri Left \\
 34 &       MCC\_R &       Cingulum\_Mid\_R &        Middle cingulate \& paracingulate gyri Right \\
 35 &       PCC\_L &      Cingulum\_Post\_L &                     Posterior cingulate gyrus Left \\
 36 &       PCC\_R &      Cingulum\_Post\_R &                    Posterior cingulate gyrus Right \\
 37 &       HIP\_L &        Hippocampus\_L &                                   Hippocampus Left \\
 38 &       HIP\_R &        Hippocampus\_R &                                  Hippocampus Right \\
 39 &       PHG\_L &    ParaHippocampal\_L &                         Parahippocampal gyrus Left \\
 40 &       PHG\_R &    ParaHippocampal\_R &                        Parahippocampal gyrus Right \\
 41 &      AMYG\_L &           Amygdala\_L &                                      Amygdala Left \\
 42 &      AMYG\_R &           Amygdala\_R &                                     Amygdala Right \\
 43 &       CAL\_L &          Calcarine\_L &      Calcarine fissure and surrounding cortex Left \\
 44 &       CAL\_R &          Calcarine\_R &     Calcarine fissure and surrounding cortex Right \\
 45 &       CUN\_L &             Cuneus\_L &                                        Cuneus Left \\
 46 &       CUN\_R &             Cuneus\_R &                                       Cuneus Right \\
 47 &      LING\_L &            Lingual\_L &                                 Lingual gyrus Left \\
 48 &      LING\_R &            Lingual\_R &                                Lingual gyrus Right \\
 49 &       SOG\_L &      Occipital\_Sup\_L &                      Superior occipital gyrus Left \\
 50 &       SOG\_R &      Occipital\_Sup\_R &                     Superior occipital gyrus Right \\
 51 &       MOG\_L &      Occipital\_Mid\_L &                        Middle occipital gyrus Left \\
 52 &       MOG\_R &      Occipital\_Mid\_R &                       Middle occipital gyrus Right \\
 53 &       IOG\_L &      Occipital\_Inf\_L &                      Inferior occipital gyrus Left \\
 54 &       IOG\_R &      Occipital\_Inf\_R &                     Inferior occipital gyrus Right \\
 55 &       FFG\_L &           Fusiform\_L &                                Fusiform gyrus Left \\
 56 &       FFG\_R &           Fusiform\_R &                               Fusiform gyrus Right \\
 57 &      PoCG\_L &        Postcentral\_L &                             Postcentral gyrus Left \\
 58 &      PoCG\_R &        Postcentral\_R &                            Postcentral gyrus Right \\
 59 &       SPG\_L &       Parietal\_Sup\_L &                       Superior parietal gyrus Left \\
 60 &       SPG\_R &       Parietal\_Sup\_R &                      Superior parietal gyrus Right \\
 61 &       IPG\_L &       Parietal\_Inf\_L & Inferior parietal gyrus, excluding supramargina... \\
 62 &       IPG\_R &       Parietal\_Inf\_R & Inferior parietal gyrus, excluding supramargina... \\
 63 &       SMG\_L &      SupraMarginal\_L &                           SupraMarginal gyrus Left \\
 64 &       SMG\_R &      SupraMarginal\_R &                          SupraMarginal gyrus Right \\
 65 &       ANG\_L &            Angular\_L &                                 Angular gyrus Left \\
 66 &       ANG\_R &            Angular\_R &                                Angular gyrus Right \\
 67 &      PCUN\_L &          Precuneus\_L &                                     Precuneus Left \\
 68 &      PCUN\_R &          Precuneus\_R &                                    Precuneus Right \\
 69 &       PCL\_L & Paracentral\_Lobule\_L &                            Paracentral lobule Left \\
 70 &       PCL\_R & Paracentral\_Lobule\_R &                           Paracentral lobule Right \\
 71 &       CAU\_L &            Caudate\_L &                               Caudate nucleus Left \\
 72 &       CAU\_R &            Caudate\_R &                              Caudate nucleus Right \\
 73 &       PUT\_L &            Putamen\_L &                   Lenticular nucleus, Putamen Left \\
 74 &       PUT\_R &            Putamen\_R &                  Lenticular nucleus, Putamen Right \\
 75 &       PAL\_L &           Pallidum\_L &                  Lenticular nucleus, Pallidum Left \\
 76 &       PAL\_R &           Pallidum\_R &                 Lenticular nucleus, Pallidum Right \\
 77 &       THA\_L &           Thalamus\_L &                                      Thalamus Left \\
 78 &       THA\_R &           Thalamus\_R &                                     Thalamus Right \\
 79 &       HES\_L &             Heschl\_L &                                Heschl's gyrus Left \\
 80 &       HES\_R &             Heschl\_R &                               Heschl's gyrus Right \\
 81 &       STG\_L &       Temporal\_Sup\_L &                       Superior temporal gyrus Left \\
 82 &       STG\_R &       Temporal\_Sup\_R &                      Superior temporal gyrus Right \\
 83 &    TPOsup\_L &  Temporal\_Pole\_Sup\_L &        Temporal pole: superior temporal gyrus Left \\
 84 &    TPOsup\_R &  Temporal\_Pole\_Sup\_R &       Temporal pole: superior temporal gyrus Right \\
 85 &       MTG\_L &       Temporal\_Mid\_L &                         Middle temporal gyrus Left \\
 86 &       MTG\_R &       Temporal\_Mid\_R &                        Middle temporal gyrus Right \\
 87 &    TPOmid\_L &  Temporal\_Pole\_Mid\_L &          Temporal pole: middle temporal gyrus Left \\
 88 &    TPOmid\_R &  Temporal\_Pole\_Mid\_R &         Temporal pole: middle temporal gyrus Right \\
 89 &       ITG\_L &       Temporal\_Inf\_L &                       Inferior temporal gyrus Left \\
 90 &       ITG\_R &       Temporal\_Inf\_R &                      Inferior temporal gyrus Right \\
 91 &   CERCRU1\_L &   Cerebellum\_Crus1\_L &               Crus I of cerebellar hemisphere Left \\
 92 &   CERCRU1\_R &   Cerebellum\_Crus1\_R &              Crus I of cerebellar hemisphere Right \\
 93 &   CERCRU2\_L &   Cerebellum\_Crus2\_L &              Crus II of cerebellar hemisphere Left \\
 94 &   CERCRU2\_R &   Cerebellum\_Crus2\_R &             Crus II of cerebellar hemisphere Right \\
 95 &      CER3\_L &       Cerebellum\_3\_L &           Lobule III of cerebellar hemisphere Left \\
 96 &      CER3\_R &       Cerebellum\_3\_R &          Lobule III of cerebellar hemisphere Right \\
 97 &    CER4\_5\_L &     Cerebellum\_4\_5\_L &         Lobule IV, V of cerebellar hemisphere Left \\
 98 &    CER4\_5\_R &     Cerebellum\_4\_5\_R &        Lobule IV, V of cerebellar hemisphere Right \\
 99 &      CER6\_L &       Cerebellum\_6\_L &            Lobule VI of cerebellar hemisphere Left \\
100 &      CER6\_R &       Cerebellum\_6\_R &           Lobule VI of cerebellar hemisphere Right \\
101 &     CER7b\_L &      Cerebellum\_7b\_L &          Lobule VIIB of cerebellar hemisphere Left \\
102 &     CER7b\_R &      Cerebellum\_7b\_R &         Lobule VIIB of cerebellar hemisphere Right \\
103 &      CER8\_L &       Cerebellum\_8\_L &          Lobule VIII of cerebellar hemisphere Left \\
104 &      CER8\_R &       Cerebellum\_8\_R &         Lobule VIII of cerebellar hemisphere Right \\
105 &      CER9\_L &       Cerebellum\_9\_L &            Lobule IX of cerebellar hemisphere Left \\
106 &      CER9\_R &       Cerebellum\_9\_R &           Lobule IX of cerebellar hemisphere Right \\
107 &     CER10\_L &      Cerebellum\_10\_L &             Lobule X of cerebellar hemisphere Left \\
108 &     CER10\_R &      Cerebellum\_10\_R &            Lobule X of cerebellar hemisphere Right \\
109 &      VER1\_2 &           Vermis\_1\_2 &                             Lobule I, II of vermis \\
110 &        VER3 &             Vermis\_3 &                               Lobule III of vermis \\
111 &      VER4\_5 &           Vermis\_4\_5 &                             Lobule IV, V of vermis \\
112 &        VER6 &             Vermis\_6 &                                Lobule VI of vermis \\
113 &        VER7 &             Vermis\_7 &                               Lobule VII of vermis \\
114 &        VER8 &             Vermis\_8 &                              Lobule VIII of vermis \\
115 &        VER9 &             Vermis\_9 &                                Lobule IX of vermis \\
116 &       VER10 &            Vermis\_10 &                                 Lobule X of vermis \\
\bottomrule
\end{longtable}

\end{document}